%% file: main.tex
\documentclass[aps,prx,twocolumn,superscriptaddress,showpacs,floatfix]{revtex4-2}
\usepackage{physics}
\usepackage{amsmath,amsthm,amssymb,amsfonts, color, comment, graphicx}
\usepackage{xcolor}
\usepackage{float}
\usepackage{dcolumn}
\usepackage{bm}
\UseRawInputEncoding 
\usepackage{dsfont}
\usepackage{hyperref}

\input{defns}

\begin{document}

\preprint{APS/123-QED}

\title{d-wave pair density wave superconductivity in a two-orbital model}
\author{Samuel Vadnais}
\email{samuel.vadnais@utoronto.ca}
\affiliation{Department of Physics, University of Toronto, 60 St. George Street, Toronto, ON, M5S 1A7 Canada}
\author{Arun Paramekanti}
\email{arun.paramekanti@utoronto.ca}
\affiliation{Department of Physics, University of Toronto, 60 St. George Street, Toronto, ON, M5S 1A7 Canada}

\date{\today}

\begin{abstract}
Motivated by exploring superconductivity in multi-orbital systems, we study a two-orbital model of spinful fermions
representing ($p_x,p_y$) or ($d_{xz}, d_{yz})$ 
orbitals on the square lattice.
For minimal interorbital $t$-$J$ or $t$-$V$ on-site interactions, a random phase approximation uncovers regimes of instability towards
incommensurate $d_{xy}$ pair density wave ($d$-PDW) superconductivity driven by interband pairing.
We study the competition of PDW order with
uniform nodal $d_{xy}$ pairing states as well as magnetic and charge density wave (CDW) instabilities. At strong coupling,
we derive an effective
hard-core Cooper pair Hamiltonian which we study using a bosonic Gutzwiller ansatz to reveal a
period-$2$ PDW over a wide range of fillings as well as a checkerboard CDW at quarter-filling. 
Our results apply to correlated multi-orbital materials with 
quasi-1D bands, Hubbard models on the square-octagon 
lattice, and atomic fermions in $p$-orbitals. Our work
highlights the role of the orbital content and multiband Fermi surfaces in stabilizing interband PDW states.
\end{abstract}

\maketitle




\section{Introduction}

The one-band Hubbard model, and effective Hamiltonians such as $t$-$J$ models,
have been extensively studied in the context of high temperature cuprate superconductors
\cite{Hubbard_review_ARCMP2022,cuprate_review_dagotto_rmp1994,cuprate_review_lee_rmp2006,cuprate_rvbreview_2004}.
These models are of broad interest in the exploration of correlated superconductivity which arises in proximity 
to Mott insulating antiferromagnets or spin density wave states
\cite{cuprate_review_lee_rmp2006}.
In the overdoped cuprates, the presence of strong quantum spin fluctuations in 
the vicinity of such ordered states could provide the `glue' for unconventional Cooper pairing and
superconductivity
\cite{scalapino_superconductivity_1999,spinflucn_abanov_epl2001,Fe_sc_chubukov2008,Fe_sc_review_chubukov2012}.

Beyond the simplest Hubbard model, which describes a single orbital at each site, 
correlated quantum materials often exhibit
multiple atomic orbitals. Such multiorbital models are important not only
as faithful microscopic models of the CuO$_2$ layers in the cuprate superconductors 
\cite{cuprate_emery1987,cuprate_emery1988,cuprate_varma1997,Cuprate4band,CuprateMai2021,Cuprateselfcons}, 
but also play a more direct role in multiband
superconductors like Sr$_2$RuO$_4$ \cite{Sr2RuO4Multi, Sr2RuO4d+ig, Sr2RuO4SpinSuscep, Sr2RuO4Suzuki2023},
and the iron-based \cite{IronSCSi2016,FeSCAFM, FeSCMagFrust,SDW_Dong_2008,SDW_RevModPhys.60.209} 
and nickel-based high temperature
superconductors \cite{nickelate_hwang_2019,nickelate_bilayer_wang_2023,nickelate_kuroki_prl2020,nickelate_wu_2019,nickelate_sahadasgupta2020,nickelate_thomale_2020,nickelate_fuchunzhang_prb2020,nickelate_werner_prb2020,nickelate_yahuizhang_2020}.
Multi-orbital models have been proposed as providing a more natural description of 
stripe and nematic orders found in correlated materials 
\cite{cuprate_Fischer2014,nematicBaek2015, nematicGlasbrenner2015, StripeWang2016, nematicWang2016}. 

Given the rich set of phenomena realized in multiorbital systems,
it is interesting to ask if multiorbital systems can
more naturally host an exotic nonzero-momentum pairing state called `pair density wave' (PDW)
order
\cite{Berg2007,PDW_annurev-conmatphys-031119-050711}
which are analogues of Larkin-Ovchinnikov \cite{Larkin1965} or Fulde-Ferrell \cite{Fulde1964} 
states but in the absence of an external Zeeman magnetic field.
Indeed, previous work has shown that multi-component
Kondo lattice systems \cite{Berg2010} and multi-orbital models \cite{pdw_paramekanti2010,pdw_jiang2023,PDW_PhysRevLett.130.026001,PDW_Setty2023,pdw_ticea2024,ALM_PDW_Fradkin2014,wang2025negativeroutepairdensity} do host PDW states, and more recent work has shown that PDW order in such
systems might be enhanced by the nontrivial quantum geometric tensor
\cite{Jiang_2023, barlas2025quantumgeometryinducedkekule,pdw_ticea2024,Lamponen2025}. 
Recent motivation to explore such exotic PDW orders comes from kagome systems \cite{wu2023sublatticeinterferencepromotespair,Yao_2025}, nickelates \cite{oh2025pairdensitywavesuperconductivityandersonstheorem},
Moir\'e materials \cite{Wu_2023,wang2026kekulesuperconductivitytwistedmagic}, 
and rhombohedral graphene \cite{Murshed_2025}.
Finally, ultracold atoms show the potential to realize multiorbital models with
tunable interactions, which makes them a formidable sandbox to study such unconventional pairing of
atomic fermions
\cite{Kohl_optical_lattice,J_rdens_2008}. 
In this paper, we study a two-orbital Hamiltonian on the square lattice, showing the emergence
of robust $d$-wave PDW states in a system with multiple Fermi surface pockets, using weak-coupling random phase
approximation and mean field theory, and bolster these results using a strong-coupling expansion at large
interaction strength.

\section{Two-orbital Model}

\begin{figure}[b]
\centering
\includegraphics[width=0.48\textwidth]{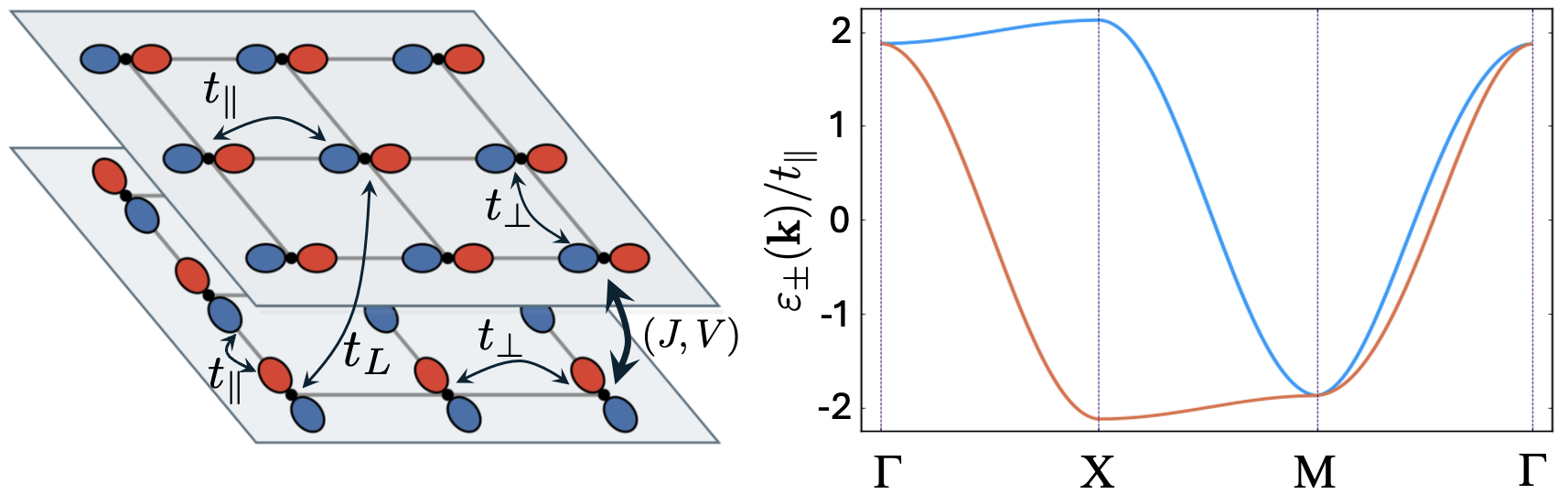}
\caption{Left: Square lattice model showing 
$(p_x,p_y)$ orbitals or $(d_{xz},d_{yz})$ orbitals. 
For visual clarity, we show the two sets of orbitals as if they are on separate layers. Intra-orbital hoppings are 
denoted by $t_\parallel, t_\perp$ and a diagonal inter-orbital hopping is denoted by $t_L$. $V,J$ respectively denote the inter-orbital 
density-density and spin exchange interactions. The model has a site-centered $C_4$ symmetry which also interchanges the
two orbitals. Right: Band dispersion of the noninteracting
two-orbital model along a high symmetry path in the Brillouin zone for $t_\perp\!=\!t_L\!=\! t_\parallel/16$.
}
\label{fig:Model}
\end{figure}

We introduce a square lattice model hosting two different orbitals at each site. These could be $(p_x,p_y)$ or $(d_{xz},d_{yz})$ orbitals which
have similar symmetry under $C_4$ lattice rotations. This lattice model is schematically represented in the top panel of  Fig.~\ref{fig:Model} for 
$(p_x,p_y)$ orbitals, and would be similar for $d$-orbitals. Representing fermion operators in these orbitals as $X,Y$, the 
Hamiltonian takes the form
\begin{eqnarray}
    H &=& H_t + H_{\rm int} \\
    H_t \!\!&=&\!\! \sum_{\bk \sigma} \left[ (\epsilon^{X}_{\bk}\!-\!\mu) X_{\bk \sigma}^\dagger X_{\bk \sigma} + (\epsilon^{Y}_{\bk}\!-\! \mu)
    Y_{\bk \sigma}^\dagger Y_{\bk \sigma} \right] \nonumber \\
    &+& \epsilon^{XY}_{\bk} \left[X_{\bk \sigma}^\dagger Y_{\bk \sigma} +  {h.c.} \right] \\
    H_{\rm int}\!\! &=&\!\! \sum_i \left[J \Vec{S}_{i}^X \cdot \Vec{S}_{i}^Y - V n_{i}^X n_{i}^Y\right]
     \label{eq:H}
\end{eqnarray}
where the interaction couples $X,Y$ at each site $i$.
The dispersions in this Hamiltonian are given by
\begin{eqnarray}
\epsilon^X_{\bk} &=& 2 t_\parallel \cos k_x - 2 t_\perp \cos k_y \\
\epsilon^Y_{\bk} &=& 2 t_\parallel \cos k_y - 2 t_\perp \cos k_x \\
\epsilon^{XY}_{\bk} &=& -4 t_L \sin k_x \sin k_y
\end{eqnarray}
We set $t_{\parallel} = 1$ to define the unit of energy, fix instances of $t_L, t_\perp$, 
and tune electron filling $0 < n < 0.5$ (where $n=0.5$ is half-filling, corresponding to 2 electrons 
per site including both spins and both orbitals) and vary the interaction strengths $V/t$ and $J/t$.
For $J=V=0$, the energy-momentum dispersion relation of the two bands are given by
\begin{eqnarray}
    \varepsilon_{\pm}(\bk)\! =\! \frac{\epsilon^X_{\bk}\!+\!\epsilon^Y_{\bk}}{2} \!\pm\! \left[\!\left(\!\frac{\epsilon^X_{\bk}\!-\!\epsilon^Y_{\bk}}{2}\!\right)^2\!\!\!\!+\!\epsilon^2_{XY}(\bk)\!\right]^{1/2}\!\!\!\!\!
\end{eqnarray}
This Hamiltonian has lattice symmetries including translation, a \(\mathcal{C}_4\) rotation around the center of each site  and time-reversal \(\mathcal{T}\). The symmetries of the Hamiltonian also include internal symmetries like \(SU(2)\) spin rotation, and a particle-hole symmetry.
Such a model arises as an effective description of the repulsive Hubbard model on the square-octagon lattice over a range of fillings \cite{Bose_Vadnais_2024}, and 
can be realized
by filling cold atoms into the $p$-orbitals of an optical lattice and having attractive local inter-orbital attraction. It may be more
generally useful as a simplified toy model of systems with quasi-one-dimensional bands and potential phonon induced inter-orbital attraction.

\subsection{Bands and Fermi surfaces versus filling}
\begin{figure}
    \centering
    \includegraphics[width=0.48\textwidth]{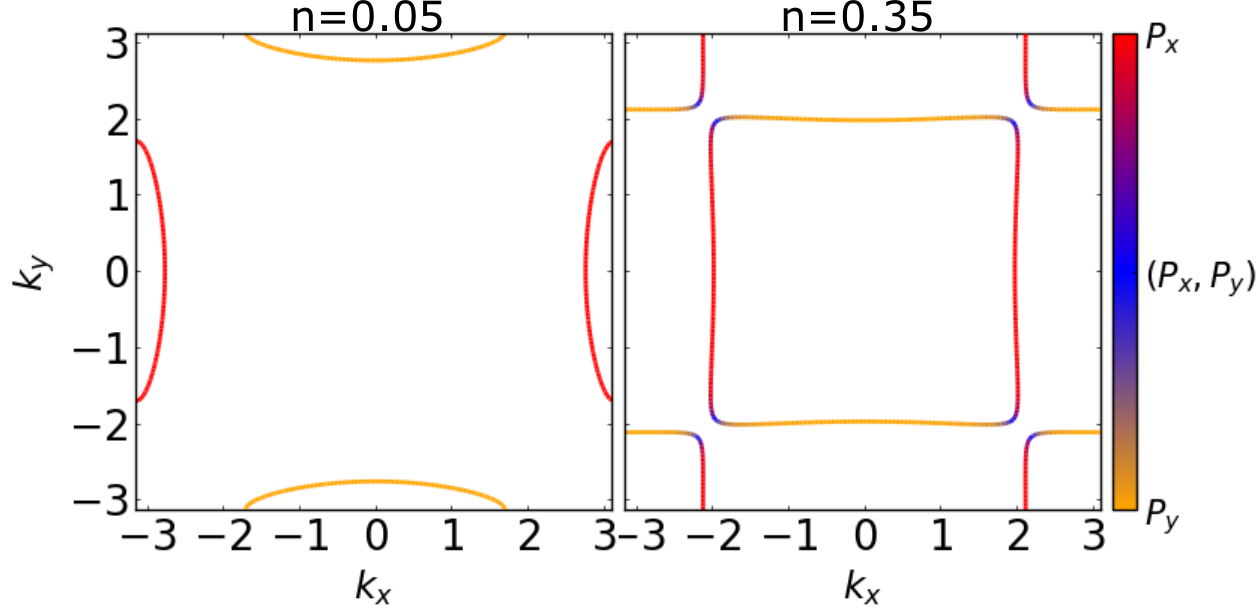}
    \caption{Fermi surfaces of two-orbital model for $t_{\perp}/t_\parallel=1/16$ at $n=0.05$ (left) and $n=0.35$ (right). The topology of the FS changes across the van Hove singularity at $n = n_{\rm vH} \approx 0.105$, which is governed by $t_{\perp},t_L$. Colors indicate the $X,Y$ orbital character of the
    FS. At low density, the two FSs with nearly pure X and pure Y character favors nonzero-momentum pairing (PDW state) driven by inter-orbital interaction.}
    \label{fig:Fermi_Surface}
\end{figure}

We begin our study of the model by inspecting the effect of $t_{\perp}$ on the Fermi surfaces. Figure \ref{fig:Fermi_Surface} shows a set of typical Fermi surfaces for the system considered at low($n=0.05$) and high ($n=0.35)$ fillings. One of its key features is its quasi-unidimensional dispersion. For a general momentum, states on the Fermi surface are mainly either $p_x$ or $p_y$ with very little mixing. Mixing of the $p_x$ and $p_y$ states is only seen along $\boldsymbol{\Gamma}$-$\boldsymbol{M}$, this is illustrated on Figure \ref{fig:Fermi_Surface}.  
As $t_{\perp}$ is introduced, the band structure of the system is distorted and its Fermi surface altered.
Its most important effect is to shift energies at the $\mathbf{M}$ and $\mathbf{X}$ points in opposite directions.
This takes the square-like Fermi surface to a four-fold symmetric set of Fermi surfaces located in the vicinity of $\bold{X}$. As $t_{\perp}$ is increased, the sufaces' curvature is accentuated and the resulting Fermi surface is displaced further from the $\boldsymbol{\Gamma}$-$\boldsymbol{M}$ line.
This results in the effective suppression of the $X_{\bold{k}}$ and $Y_{-\bold{k}}$ momentum pairing, in favor of finite-momentum pairing. This is expected to impact pairing as $p_x$/$p_y$ states are no longer available for zero-momentum pairing. For intermediate filling, $t_{\perp}$ has little to no effect on the overall topology of the Fermi surface.


\subsection{RPA Susceptibilities}

The leading instabilities of the normal state in the weak coupling regime can be evaluated by considering the orbital 
resolved magnetic, charge, and pairing susceptibilities. Below, we calculate these within RPA to identify possible
broken symmetry states which can arise from the interaction terms in the Hamiltonian.

For the particle-hole instabilities, we use the result for the fermion bubble,
\begin{align}
    \chi^{0}_{abcd}&(\bQ,i\W_n) \!=\!  -\frac{1}{N} \! \sum_{\bk,nm} \! \sum_{i\W_n} \! \frac{n_F(\varepsilon_n(\bk)) - n_F(\varepsilon_m(\bk\!+\!\bQ))}{\varepsilon_n(\bk) - \varepsilon_m(\bk\!+\!\bQ) + i\W_n}  \nonumber \\  \times &\left[u_{bn}(\bk)u_{cn}^*(\bk) u_{dm}(\bk\!+\!\bQ)u_{am}^*(\bk\!+\!\bQ)\right] \d_{\b\g}\d_{\a\d},
\end{align}
to express the charge and spin susceptibility tensors as:
\begin{align}
    \chi^{0z}_{abcd}(\bQ,i\W_n) &=  \s^{\a\b,z}_{ab}  \chi^{0,\a\b\g\d}_{abcd}(\bQ,i\W_n) \s^{\g\d,z}_{cd} \nonumber \\
     &=  \frac{1}{2} \chi^{0}_{abcd}(\bQ,i\W_n)
\end{align}
\begin{align}
  \chi^{0c}_{abcd}(\bQ,i\W_n) & =  \s^{\a\b,0}_{ab}  \chi^{0,\a\b\g\d}_{abcd}(\bQ,i\W_n) \s^{\g\d,0}_{cd}\nonumber \\
  &=  2 \chi^{0}_{abcd}(\bQ,i\W_n)
\end{align}
Here $ \chi^{0}_{abcd}(\bQ,i\W_n)$ is the orbitally resolved bare particle-hole bubble, at momentum $\bQ$ and frequency $i\Omega_n$, and latin/greek indices are orbital/spin labels. 

\begin{figure}[hbt]
    \centering
    \includegraphics[width=0.98\linewidth]{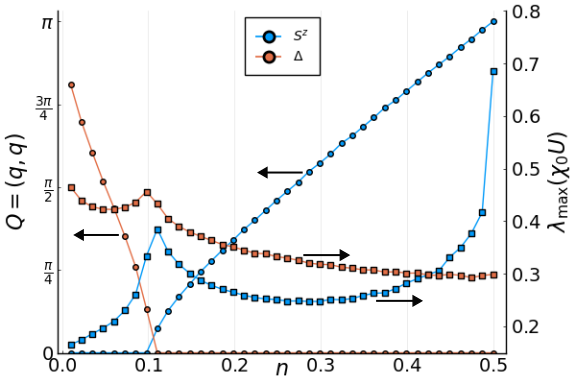}
    \caption{Evolution of the spin and pairing $\lambda_{\text{max}}(\chi_0 U_{\text{eff}})$ (square markers,right axis) and corresponding wave-vector instability(circle markers,left axis) $\bQ = (q,q)$ vs filling, at $t_{\perp}=t_\parallel/16$ and $V=0$. Finite-momentum pairing is dominant at fillings $n < n_{\text{vH}}$, while the spin channel exhibits finite-momentum instabilities for $n > n_{\text{vH}}$.}
    \label{fig:Susceptibility_Q}
\end{figure}

For pairing, it is natural to consider orbital-triplet spin-singlet (OTSS) pairing states, given the form and sign of the local
interaction. We note that this pairing order parameter transforms as a $d_{xy}$ SC. With this consideration, OTSS pairing susceptibility reads:
\begin{align}
   &\chi^{0p}(\bQ,i\Omega_n) = \sum_{\bk} \sum_{n,m}  \frac{1-n_F(\varepsilon_n(-\bk+\bQ)) - n_F(\varepsilon_m(\bk))}{\varepsilon_n(-\bk+\bQ) + \varepsilon_{m}(\bk) - i\Omega_n} \nonumber\\ & \times \bigg[u_{Yn}(-\bk+\bQ) u_{Xn}^*(-\bk+\bQ) u_{Xm}(\bk) u_{Ym}^*(\bk) \nonumber \\ 
   &+|u_{Yn}(-\bk+\bQ)|^2 |u_{Xm}(\bk) |^2 \bigg] 
\end{align}
In the random phase approximation, an instability in any of these channels will be marked by the divergence of the RPA susceptibility:
\begin{align}
    \chi^{\tau,RPA} = & (1+\chi^{0,\tau} U_{\text{eff}}^\tau)^{-1} \chi^{0,\tau}
\end{align}
with $U_{\text{eff}}^\tau$ the effective interaction tensor in orbital space for channel $\tau$. This divergence occurs whenever the largest eigenvalue $\lambda_{\text{max}}( \chi^{0,\tau}(\bQ) U_{\text{eff}}^\tau) = 1$, and the nature of the instability in orbital space is revealed by the associated eigenvector in the orbital basis. The effective interactions in each of these channels are derived by manipulating the  interaction.
\begin{align}
    U_{abcd}^z &= 
  \begin{cases}
    J & a=b,c=d,a\neq c\\
    0 &\text{otherwise}\end{cases}\\
    U^p &=  - \frac{3}{4} J - V\\
    U_{abcd}^c &= 
  \begin{cases}
    -V & a=b,c=d,a\neq c \\
    0 &\text{otherwise}\end{cases}
\end{align}
Figure \ref{fig:Susceptibility_Q} presents the evolution of these instabilities for $V=0$; since the model is particle-hole symmetric, we restrict
attention to $0 \leq n \leq 0.5$.
Broadly, we find
singlet PDW $d_{xy}$ superconductivity with incommensurate Cooper pair momentum $\bQ$ at low densities $n < n_{\rm vH}$, uniform $d_{xy}$ SC for
densities $n_{\rm vH} < n \lesssim 0.45$, and incommensurate magnetic order for $n \gtrsim 0.45$.

\begin{figure}
 \centering
\includegraphics[width=0.98\linewidth]{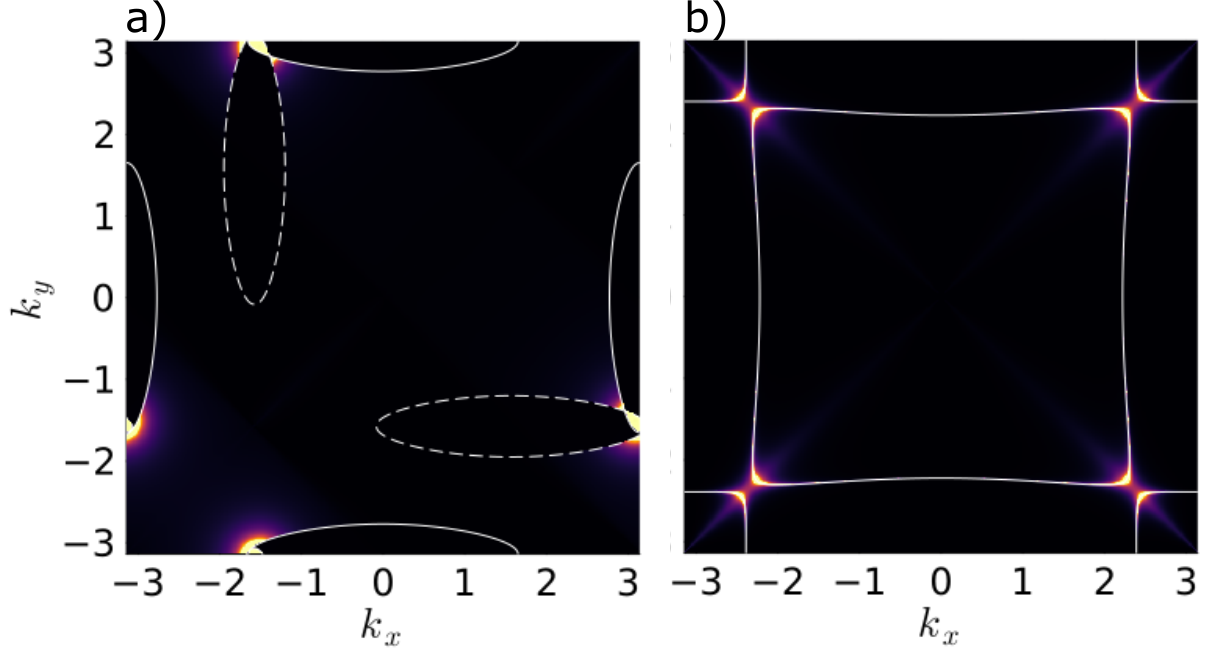}
    \caption{a) Pairing heatmap of the IC PDW($Q \approx \frac{\pi}{2}$) superconductor at filling $n=0.05$, and b) Uniform SC at filling $n=0.27$. Fermi surfaces are drawn with white lines, while the dotted lines represent the shifted $-\bk+\bQ$ Fermi surfaces.}
    \label{fig:PairingOverlap}
\end{figure}

Based on the Fermi surface topology and orbital character discussed earlier, the dominant pairing instability at low fillings is towards a
nonzero momentum 
$\bQ=(q,q)$ Cooper pairing. The wavevector $\bQ$ vector continuously shifts from $\bQ=(\pi,\pi)$ at very low fillings to 
$\bQ=(0,0)$ at 
$n_{\rm vH} \approx 0.1$. This $\bQ$ vector instability selection can be understood as being the momentum vector maximizing the overlap of 
$X_\bk$ and $Y_{-\bk+\bQ}$ states, as illustrated in Fig.~\ref{fig:PairingOverlap}(a). 
This finite-momentum pairing is driven by $X$-$Y$ orbital electron pairing which occurs between different $C_4$ symmetry related bands.
At the van-Hove singularity, 
the Fermi surface reconstructs, and beyond this point we find sections of the Fermi surface along the 
$\bold{\Gamma}-\bold{M}$ direction with strong hybridization of 
$p_x$ and $p_y$ orbitals. The pairing between these corner regions on the Fermi surfaces, shown in Fig.~\ref{fig:PairingOverlap}(b),
leads to the stabilization of a 
uniform $Q=(0,0)$ SC. In this regime, we find dominant intraband pairing on both the Fermi surface sheets. 

We note
that the uniform SC features a local singlet order parameter 
$\Delta=\langle X^\dg_\upa Y^\dg_\dna - X^\dg_\dna Y^\dg_\upa\rangle$. Under $C_4$ rotations, $X \to Y$, $Y \to -X$, so that
$\Delta \to -\Delta$. Thus the uniform SC is a $d_{xy}$ superconductor. As we discuss below, this SC features point nodes on
the multiband Fermi surfaces for weak to moderate couplings, but has a full spectral gap at strong coupling. The modulated
$\bQ$-PDW state which arises from this corresponds to a $d$-PDW state.

Over most of the density range, we find that the magnetic susceptibility remains finite when the pairing susceptibility diverges, 
so the magnetic instability is not the primary instability for $n\lesssim 0.45$. Beyond this point, we find a dominant tendencey
towards an incommensurate magnetic order with $\bQ=(\pi,\pi)$ at half-filling. 
The eigenvector corresponding to this instability is of the form: $(1,0,0,-1)^T$ suggesting a magnetically compensated state 
where each orbital forms $\bQ$ magnetic order and the two orbitals have opposite magnetization.
In contrast to the effect of $J$, an inter-orbital density-density interaction $V$ would instead promote an instability in the charge 
channel for $n \gtrsim 0.45$ while suppressing magnetic order. However, $V$ also promotes pairing in the OTSS channel, so we
expect that having both $J,V$ would favor pairing over most of the density regime within RPA.

Although the 
Stoner criterion can identify the leading linear instability of the system, the presence of multiple coexisting orders are still possible 
in the eventuality that multiple susceptibilities are diverging for a given interaction. Studying this interplay needs more sophisticated
methods such as the functional renormalization group method \cite{liu2026genuinepairdensitywave} which is beyond the scope of this paper. Below we carry out a mean field study of
this model, solving the nonlinear gap equation, in order to explore possible ground states and coexisting orders
as a function of interaction strength.

\section{Mean field theory}

To go beyond the RPA instability and explore the interaction dependence of PDW orders in the phase diagram, we resort to two complementary
mean field approaches. If we assume that the PDW order is of the Fulde-Ferrell type, with Cooper pairs carrying momentum $\bQ$, we can 
recast the mean field theory in momentum space and solve self-consistently for the order parameters. By contrast, if the PDW is of the
Larkin-Ovchinnikov type, we need to incorporate modulations of the pair amplitude and density which is easier to explore within a real
space Bogoliubov-deGennes mean field approach. 

\subsection{Momentum space mean field theory}
Assuming a Fulde-Ferrell type of OTSS pairing, the PDW order parameter can be written:
\begin{align}
    \Delta_0 = \sum_\bk \expval{X_{\bk\ua}^\dagger Y_{-\bk+\bQ \da}^\dagger - X_{\bk\da}^\dagger Y_{-\bk+\bQ \ua}^\dagger}
\end{align}
For $\bQ=0$, under $C_4$ rotations $X \to Y$, $Y \to -X$, such that the order parameter changes sign. Thus the uniform SC is a $d_{xy}$ superconductor. The modulated
$\bQ$-PDW state which arises from this could be termed a $d$-PDW state in the sense of having a local on-site order parameter which
changes sign under $C_4$ rotations, but the full superconducting state will
mix different irreducible representations of the square lattice point group symmetry.

The mean-field Hamiltonian can be written as 
\begin{align}
    H &= \sum_\bk  \Psi_{\bk}^\dagger H_{\rm MF}(\bk) \Psi_{\bk}^\pdg
    \nonumber \\
   H_{\rm MF}(\bk) &=
\begin{pmatrix}
    \xi^X_{\bk} & \e_{\bk}^{XY} & 0 &U_{\text{eff}}\D_0 \\
    \e_{\bk}^{XY} & \xi^Y_{\bk} & U_{\text{eff}}\D_0 & 0 \\
    0 & U_{\text{eff}}\D_0^* & -\xi^X_{-\bk+\bQ} & -\e_{-\bk+\bQ}^{XY} \\
    U_{\text{eff}}\D_0^* & 0 & -\e_{-\bk+\bQ}^{XY} & -\xi^Y_{-\bk+\bQ}
\end{pmatrix}
\end{align}
with $\Psi^\dg_{\bk} \!=\! \left(
X^\dg_{\bk\ua},
Y^\dg_{\bk\ua},
X_{-\bk+\bQ\da}^\pdg,
Y_{-\bk+\bQ\da}^\pdg\right)$,
$\xi^{X}_{\bk}\!=\! \e_{\bk}^{X} \!-\! \mu$, and $\xi^{Y}_{\bk}\!=\! \e_{\bk}^{Y} \!-\! \mu$.
This Hamiltonian is iteratively solved at each filling and interaction strength to determine the chemical potential and the 
SC pair amplitude $\Delta_0$. For the ground state, we work at temperature $T/t_\parallel=10^{-3}$, and solve for the ground state
for every momentum vector $\bQ$, until a $10^{-4}$ convergence criteria is reached. The energy is then evaluated for each 
solution yielding the minimum energy superconducting phases shown in Figure \ref{fig:tpp_0.06_PhaseDiagram}. Since the 
magnetic phase is not the primary concern of the current work, we infer this phase from the RPA magnetic
susceptibility derived previously, but we schematically sketch the shape of the magnetic phase boundary to highlight the fact
that the strong coupling limit which leads to
formation of $X$-$Y$ interobital singlet formation which will eventually suppress the magnetic order at large $J$.
\\

\subsection{Real space mean field theory}

To solve the model in real space, we decompose the interaction terms into the pairing and particle-hole channels by incorporating
suitable Weiss fields. 
The generalized mean-field decoupled Hamiltonian considered in real space is the following:
\begin{eqnarray}
\!\!\!\!\!    H^{\rm MF}_{ij} \!&=&\! \left( \! H_{\rm loc} \!+\! H_{\rm pair}\right)\,,\\
\!\!\!\!\!    H_{\rm loc} \!&=&\! c_{i\a}^\dagger \left(\! -\mu\d^{\a\b} \!+\! \frac{J_{ij}}{4} \s_{a}^{\a\b}\s_{a}^{\g\d}M_j^{\g \d} \! \right) 
c_{i\b} \!+\! {i \leftrightarrow j} \\
\!\!\!\!\!    H_{\rm pair} \!&=&\! c_{i\a}^\dagger \left(\frac{J_{ij}}{4} \s_{a}^{\a\g}\s_{a}^{\b\d}\Delta_{ji}^{\d \g} \right) 
c^{\dagger}_{j\b} \!+\! {\rm h.c.}
\end{eqnarray}
where $\D_{ij}^{\a \b} \!=\! \expval{c_{i\a}c_{j\b}}$,
$M_{i}^{\a \b} \!=\! \expval{c_{i\a}^\dagger c_{i\b}}$. Solving for these order parameters within this real space framework on a unit-cell of size commensurate with the wave-vector allows to probe wether the SC state is of the Fulde-Ferrel and/or Larkin-Ovchnikov type at ordering wave-vectors commensurate with the unit-cell. 
We solve the resulting mean field equations in real space with suitable unit-cell size and periodizing in
momentum space to capture the 
expected instability. For $2 \times 2$ and $4\times 4$ unit-cells, we use momentum grids $100\times 100$ and $25 \times 25$
respectively which allows reasonable k-space resolution and computing time.
All channels are solved self-consistently, with a convergence tolerance $\sim 10^{-4}$ for each mean-field order parameter. We find that
the instability is towards a Fulde-Ferrell PDW state, with negligible density modulations.




\par

\par

\subsection{Phase diagram}

\begin{figure}[b]
    \centering
    \includegraphics[width=0.95\linewidth]{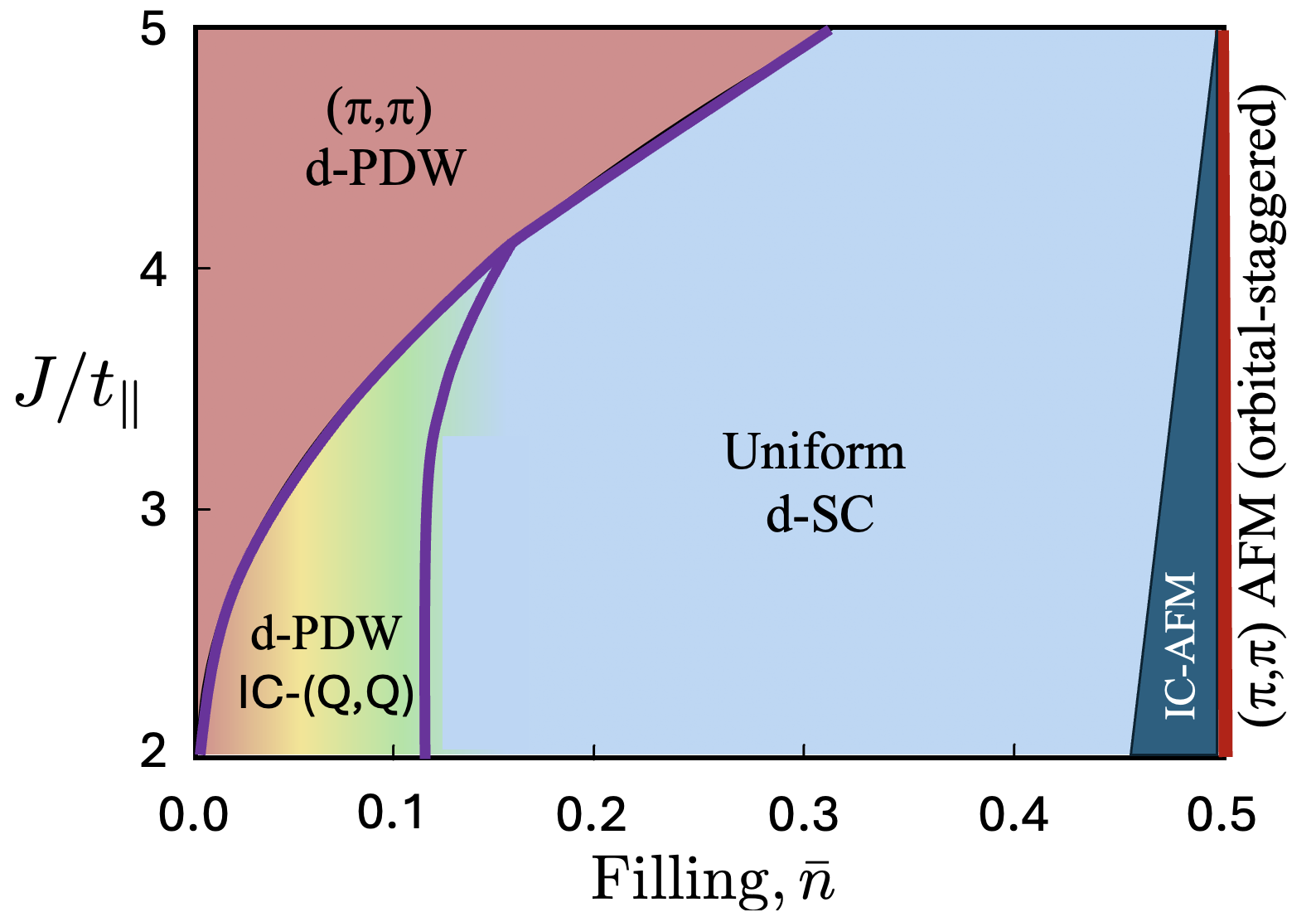}
    \caption{Qualitative phase diagram as a function of magnetic exchange $J$ and filling $n$ obtained from a combination of momentum 
    space mean-field theory calculations and Bogoliubov-deGennes computations on periodized systems using $2\times 2$ and $4\times 4$
     unit cell sizes.}   
    \label{fig:tpp_0.06_PhaseDiagram}
\end{figure}

Based on the results from both the above approaches we arrive at the phase diagram shown in
Fig.~\ref{fig:tpp_0.06_PhaseDiagram}. For smaller values of $J/t_\parallel$, the mean field phases we find (incommensurate PDW, uniform SC, and
incommensurate magnetic orders) are consistent with our RPA calculations. With increasing interaction strength, we find that the incommensurate
PDW order as well as the uniform SC gives way to a commensurate period-$2$ PDW with $\bQ=(\pi,\pi)$. We explain this dominance of the $(\pi,\pi)$
PDW order below using a strong coupling expansion for $J/t_\parallel \gg 1$. Similarly, at half-filling, the AFM order found at weak coupling
eventually gives way to a featureless insulating state formed by having gapped inter-orbital singlets at each site.

\subsection{Bogoliubov quasiparticle spectrum}
\begin{figure}[t]
    \centering
\includegraphics[width=0.95\linewidth]{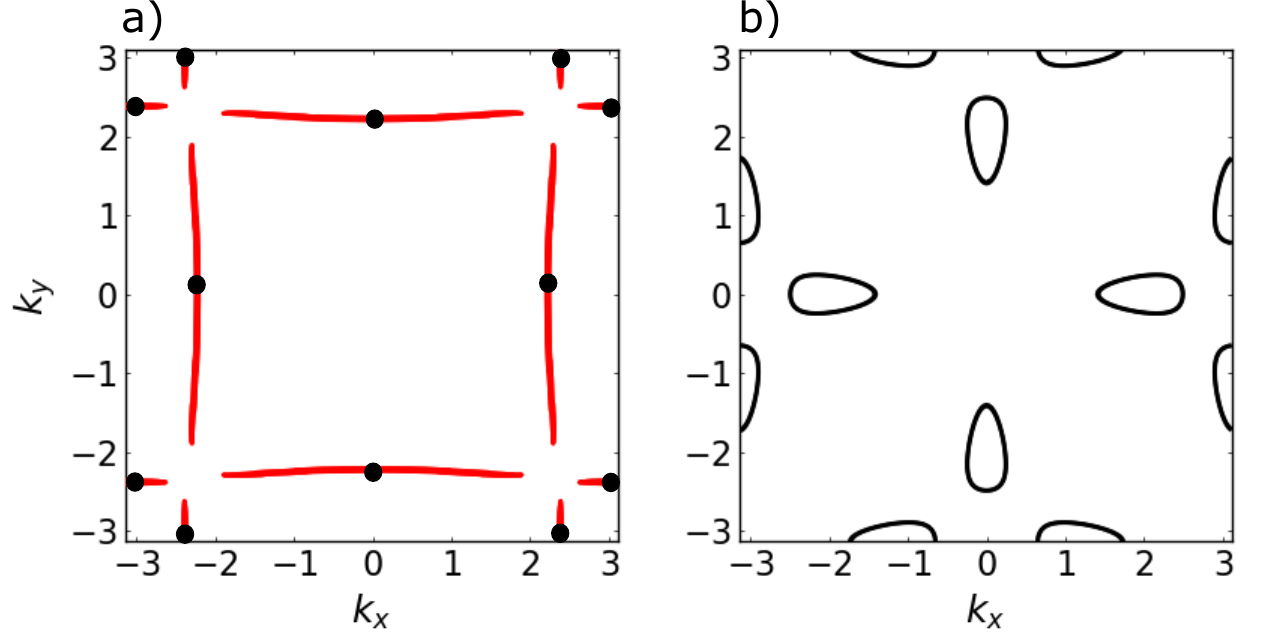}
    \caption{BdG excitation spectrum showing (a) symmetry enforced gapless d-wave nodes (black dots) and low energy 
    excitations (red contours) 
    for $n=0.27$, in the $Q=(0,0)$ uniform $d_{xy}$ superconductor at J=$4.0 t_\parallel$, and (b) gapless Fermi contours of
    Bogoliubov quasiparticles
    for $n=0.05$ in the $Q=(\pi,\pi)$ PDW state at $J=3.25 t_\parallel$.
    }
    \label{fig:BdGSpectrum}
\end{figure}

In this section we further look into the Bogoliubov quasiparticle spectrum for the uniform superconducting state and the PDW state. 
For simplicity, we restrict our discussion of the PDW order to the $(\pi,\pi)$ PDW state.
As seen from Fig.~\ref{fig:BdGSpectrum}(a), the quasiparticle spectrum shows the contours of zero and low energy Bogoliubov
excitations. We find zero gap $d$-wave nodes in the spectrum (black dots) along with low energy
excitations which at low nonzero energy form highly anisotropic (red) contours in momentum space. The location of the nodes is consistent
with symmetry; we expect the $d_{xy}$ order to have a gap of the form $\sin k_x \sin k_y$ which should vanish along lines
with $k_x=0,\pi$ and $k_y=0,\pi$, so the nodes would appear where these lines intersect the multiband Fermi surfaces leading to 
the indicated point nodes. For the $(\pi,\pi)$ PDW state, we instead find closed contours of gapless excitations resembling a
Bogoliubov Fermi surface with multiple pockets.

\section{Strong coupling description of the PDW}
In this section we derive strong coupling theory explaining the emergence of a $(\pi,\pi)$ PDW state observed in our mean field phase diagram. 
For convenience, we work with $H_{\rm int}\!\!=\!\! \sum_i J \Vec{S}_{i}^X \cdot \Vec{S}_{i}^Y$, setting $V=0$, but our results easily
generalize in the presence of $V$ with the simple replacement $3J/4 \to 3 J/4+V$.
We define local bosonic Cooper pair creation operators via
\begin{align}
    b_i^\dagger = X^\dagger_{i\uparrow}Y^\dagger_{i\downarrow} - X^\dagger_{i\downarrow}Y^\dagger_{i\uparrow}
\end{align}
The low energy states at strong coupling are spanned by configurations $\ket{0}$ or $\ket{S_i} = b_i^\dagger\ket{0}$
at each site $i$, so that each site is either empty or has an inter-orbital Cooper pair. 
In the absence of Cooper pair hopping, each singlet pair will have energy $e_0= E_0/N_{\rm pair} = -3 J/4$,
where $N_{\rm pair}/N = 2 n$, $N$ denotes the number of sites on the lattice and $n$ is the
fermion filling.

\subsection{Strong coupling perturbation theory}

Following the method in $\cite{takahashi_half_filled_1977}$, the perturbative expansion can be written up to second order in hopping as
\begin{align*}
    H_{\rm eff} & = H_{\rm int} + H_2\\
    H_2  &= P_0 H_t S H_t P_0
\end{align*}
where $H_t$ is the fermion hopping Hamiltonian in Eq.~\ref{eq:H}, $P_0$ is a projector to the strong coupling configurations
above, and 
\begin{eqnarray}
    S = \frac{(1-P_0)}{E_0 - H_{\text{int}}}.
\end{eqnarray}
To second order in this perturbative expansion, the effective low-energy Hamiltonian in the strong coupling limit is the following:
\begin{equation}
    H_{\text{eff}} \!=\!  \sum_{ij} t_{ij}^{\text{eff}} b_i^\dagger b_j \!+\!\! \sum_{ij} \! W_{ij} n^b_i (1\!-\! n^b_j) \!+\! (e_0\!-\!\mu_{\rm eff}) 
   \! \sum_i\! b_{i}^\dagger b_i
\end{equation}
where the Cooper pair hopping terms
\begin{eqnarray}
    t^{\rm eff}_{\la ij \ra } &=& \frac{8t_{\parallel} t_{\perp}}{3J};~~t^{\rm eff}_{\la\la ij \ra\ra} = -\frac{8 t_L^2}{3J}
\end{eqnarray}
correspond to nearest-neighbor and next-nearest neighbor hopping terms respectively, while the Cooper pair density $n^b_i$
enters in the
interaction terms, with
\begin{eqnarray}
   \!\!\! \!\!\!\!\! W_{\la ij\ra} \!\!&\equiv&\!\! W_1 \!=\! -\frac{4}{3J}(t_{\parallel}^2 \!+\! t_{\perp}^2);~ W_{\la\la ij \ra\ra} \!\equiv\! W_2 \!=\! 
    -\frac{8}{3J} t_L^2
\end{eqnarray}
corresponding respectively to nearest-neighbor $\la ij\ra$ and next-nearest neighbor $\la\la ij\ra\ra$ pairs of sites.
These $W$ terms represent the energy lowering from the virtual back and forth hopping of fermions between neighboring
pairs of filled and
unfilled sites. Ignoring 
the hard-core constraint of one Cooper pair per site and the short-distance Cooper pair interactions encoded by $W$,
the effective Cooper pair hopping Hamiltonian can be written in momentum space as
\begin{eqnarray}
   \! H_{\rm eff} \!\!&=&\!\! \sum_\bQ (e_0 +E^b_{\bQ}-\mu_{\rm eff}) b_\bQ^\dagger b^\pdg_\bQ
   \end{eqnarray}
   with
   \begin{eqnarray}
        \!\!\!\!\!\!\! \!\!\! E^b_{\bQ}\!\!&=&\!\! \frac{16 t_{\parallel} t_{\perp}}{3J} (\cos Q_x  \!+\! \cos Q_y) \!-\! \frac{32 t_L^2}{3J} \cos{Q_x}\cos{Q_y}.
   \end{eqnarray}
We find that $E^b_{\bQ}$ has a minimum at $\bQ=(\pi,\pi)$ due to the negative sign of the Josephson coupling (i.e., Cooper pairing hopping)
since the $X,Y$ electrons hop with the opposite sign when moving to the nearest neighbor site.
 This explains why we find the $\bQ=(\pi,\pi)$ PDW to be favored in the limit of strong coupling. 
 As coupling is slowly tuned down, higher order perturbative terms in the fermion hopping 
 are expected become more important and the bosonic dispersion minimum 
 is expected to be shifted. This could explain the change of $\bQ$ to incommensurate values in the phase diagram as interaction is decreased
 from the strong coupling limit.

 \subsection{Gutzwiller ansatz}
 
To go beyond this non-interacting Cooper pair limit, we use a mean field Gutzwiller ansatz, where we write the boson
wavefunction as a direct product of boson wavefunctions at each site which respect the no-double-occupancy constraint. In this case,
we can represent the boson wavefunction as
\begin{eqnarray}
    |\psi_b\ra = \prod_{\otimes i} \left[\cos\theta_i e^{-i \phi_i} |n^b_i=1\rangle + \sin\theta_i |n^b_i=0\rangle \right]
\end{eqnarray}
where $|n^b_i=0\ra,|n^b_i=1\ra$ respectively represent states with zero and one singlet $XY$ Cooper pair at site $i$.
Using this, we find the Gutzwiller ansatz energy
\begin{eqnarray}
    E^b_{\rm Gutz} &=& -\mu \sum_i \cos^2 \theta_i + W_1 
    \sum_{\langle ij\rangle} \cos^2 \theta_i \cos^2 \theta_j \nonumber \\
    &+& W_2 \sum_{\langle\langle ij\rangle\rangle} \cos^2 \theta_i \cos^2 \theta_j
    \nonumber \\
    &-& \frac{4 t_\parallel t_\perp}{3 J} \sum_{\langle ij\rangle} \sin 2\theta_i \sin 2\theta_j 
    \cos(\phi_i-\phi_j) \nonumber \\
   & - &\frac{4 t_L^2}{3 J} \sum_{\langle\langle ij\rangle\rangle} \sin 2\theta_i \sin 2\theta_j 
    \cos(\phi_i-\phi_j)
\end{eqnarray}
Numerically minimizing this energy function, we find a $(\pi,\pi)$ PDW over most of the filling range. In addition,
at half-filling where $n=0.5$ (or $n^b=1$ Cooper pair per site), we find a trivial Mott insulator, while at
quarter-filling where $n=0.25$ (or $n^b=0.5$ Cooper pair per site) we find an insulating checkerboard
charge density wave state. This strong coupling expansion could be extended by incorporating boson fluctuations
beyond the Gutzwiller approximation; we defer this to future work.

\section{Discussion}
In summary, we have studied the competition between zero and finite-momentum $d$-wave pairing in a multiorbital model. We have  described how the 
PDW state can be enhanced at the mean-field level with purely local interactions in the presence of multiple bands at the Fermi level. 
This is discussed here using a two-orbital model of spinful fermions representing $(p_x/d_{xz},p_y/d_{yz})$ orbitals, systems of potential interest in
solid state and cold atom settings.
The corresponding phase diagram exhibits both an incommensurate d-wave pair density wave state at low filling and a uniform superconducting state as 
filling is tuned. We highlight the importance of Bloch functions form factors and Fermi surface topology in the stabilization of PDW    
states in multiorbital systems. Finally, we have presented a strong-coupling effective theory for which a $\bQ=(\pi,\pi)$ pair density 
wave emerges naturally.

\acknowledgments
We thank Srinivas Raghu and Yafis Barlas for helpful discussions.
This research was funded by the Natural Sciences and Engineering Council of Canada (NSERC) via Discovery Grant RGPIN-2026-04578. 
SV acknowledges
support through an Ontario Graduate Scholarship (OGS) award. Numerical computations were performed on the Trillium supercomputer at the SciNet HPC Consortium and the Digital Research Alliance of Canada.

\bibliography{PDWpaper}

\appendix 
\onecolumngrid
\clearpage
\section{Weak Coupling}
\subsection{Charge/Spin Susceptibility}

We start by defining the generalized particle-hole bubble:
\begin{align}
    \chi^{0,\a\b\g\d}_{abcd}(r,r',\tau) = & T_\t \expval{c_{a\a}^\dagger(r,\tau)  c_{b\b}(r,\tau) c_{c\g}^\dagger(r',0)  c_{d\d}(r',0)}_c \\
    \chi^{0,\a\b\g\d}_{abcd}(\bQ,\tau) = & \sum_\bk \frac{-1}{N} G_{bc}(\bk,\tau)G_{da}(\bk+\bQ,-\tau) \d_{\b\g}\d_{\a\d} \\
   \chi^{0,\a\b\g\d}_{abcd}(\bQ,i\W_n) = & \frac{-1}{N\b} \sum_{k,nm} \sum_{i\w_n}\left[U_{bn}(\bk)U_{cn}^*(k) U_{dm}(\bk+\bQ)U_{am}^*(\bk+\bQ)\right] G_n^{\b\g}(\bk,i\w_n)G_m^{\a\d}(\bk+\bQ,i\w_n + i\W_n) \d_{\b\g}\d_{\a\d} \\ 
  \chi^{0}_{abcd}(\bQ,i\W_n) \d_{\b\g}\d_{\a\d}  = & \frac{-1}{N} \sum_{\bk,nm}  \sum_{i\w_n} \left[U_{bn}(\bk)U_{cn}^*(\bk) U_{dm}(\bk+\bQ)U_{am}^*(\bk+\bQ)\right]\frac{n_F(\e_n(\bk)) - n_F(\e_m(\bk+\bQ))}{\e_n(\bk) - \e_m(\bk+\bQ) + i\W_n} 
\end{align}
Where greek and latin indices label spin and orbital degrees of freedom respectively.
From this tensor, we derive the magnetic susceptibility $\chi^{0z}_{abcd}(Q,i\W_n)$ and the charge susceptibility $\chi^{0c}_{abcd}(\bQ,i\W_n)$:
\begin{align}
   \chi^{0z}_{abcd}(\bQ,i\W_n) = & \sum_{\a\b\g\d}\frac{1}{4} \s^{\a\b,z}_{ab} \s^{\g\d,z}_{cd} \chi^{0}_{abcd}(\bQ,i\W_n) \d_{\b\g}\d_{\a\d} \nonumber \\
   = & \frac{1}{4} \Tr(\s^z)^2 \chi^{0}_{abcd}(\bQ,i\W_n) \nonumber \\
  = & \frac{1}{2}  \chi^{0}_{abcd}(\bQ,i\W_n) \\
  \chi^{0c}_{abcd}(\bQ,i\W_n) = & \s^{\a\b,0}_{ab} \s^{\g\d,0}_{cd} \chi^{0,\a\b\g\d}_{abcd}(Q,i\W_n) \nonumber \\
  = & 2 \chi^{0}_{abcd}(\bQ,i\W_n)
\end{align}
\subsection{Singlet Pairing Susceptibility}
From the sign and overall form of the interactions considered, we derive the spin singlet orbital triplet pairing susceptibility. We define the spin singlet pairing operator and the pairing susceptibility:
\begin{equation}
    \Delta^\dagger(\bQ) = \frac{1}{\sqrt{2}}\sum_\bk (X_{\bk\ua}^\dagger Y_{-\bk+\bQ\da}^\dagger - X_{\bk\da}^\dagger Y_{-\bk+\bQ\ua}^\dagger)
\end{equation}
\begin{align*}
    \chi^p(\bQ,i\Omega_n) = & \int_0^\beta  e^{i\Omega_n \tau} d\tau \expval{\Delta(\bQ,\tau)\Delta^\dagger(\bQ,0)}\\
    = & \frac{1}{2}\int_0^\beta  e^{i\Omega_n \tau}  d\tau \sum_{\bk \mathbf{k'}} \left( T_\tau \expval{Y_{-\bk+\bQ\da}(\tau) Y^\dagger_{-\mathbf{k'}+\bQ \da}X_{\bk\ua}(\tau) X^\dagger_{\mathbf{k'}\ua}} - T_\tau\expval{Y_{-\bk+\bQ\da}(\tau) Y^\dagger_{-\mathbf{k'}+\bQ\ua} X_{\bk\ua}(\tau) X^\dagger_{\mathbf{k'}\da}} \right.  \\ 
    - & \left. T_\tau \expval{Y_{-\bk+\bQ\ua}(\tau) Y^\dagger_{-\mathbf{k'}+\bQ\da} X_{\bk\da}(\tau) X^\dagger_{\mathbf{k'}\ua}} + T_\tau \expval{Y_{-\bk+\bQ\ua}(\tau) Y^\dagger_{-\mathbf{k'}+\bQ\ua} X_{k\da}(\tau) X^\dagger_{\mathbf{k'}\da}} \right) \\
    = & \frac{1}{2} \int_0^\beta  e^{i\Omega_n \tau} d\tau  \sum_\sigma \sum_{\bk\mathbf{k'}} T_\tau \left( \expval{Y_{-\bk+\bQ\sigma}(\tau) Y^\dagger_{-\mathbf{k'}+\bQ \sigma}X_{\bk \bar{\sigma}}(\tau) X^\dagger_{\mathbf{k'}\bar{\sigma}}} + \expval{Y_{-\bk+\bQ\sigma}(\tau) X^\dagger_{\mathbf{k'}\sigma} X_{\bk\bar{\sigma}}(\tau) Y^\dagger_{-\mathbf{k'}+\bQ \bar{\sigma}}} \right) \\
    = & \frac{1}{\beta}\sum_{i\nu_n} \sum_{\bk} \sum_{n,m'}  \left[ |U_{Yn}(-\bk+\bQ)|^2 |U_{Xm'}(\bk) |^2 + U_{Yn}(-\bk+\bQ) U_{Xn}^*(-\bk+\bQ) U_{Xm'}(\bk) U_{Ym'}^*(\bk)\right] \\
     & G^n(-\bk+\bQ,i\nu_n)  G^{m'}(\bk,i\Omega_n-i\nu_n) \\
     = & \sum_{\bk} \sum_{n,m'}  \left[ |U_{Yn}(-\bk+\bQ)|^2 |U_{Xm'}(\bk) |^2 + U_{Yn}(-\bk+\bQ) U_{Xn}^*(-\bk+\bQ) U_{Xm'}(\bk) U_{Ym'}^*(\bk)\right] \\
     & \frac{1-n_F(\epsilon_n(-\bk+\bQ)) - n_F(\epsilon_m'(\bk))}{\epsilon_n(-\bk+\bQ) + \epsilon_{m'}(\bk) - i\Omega_n}
\end{align*}

\subsection{RPA effective interactions}
The interactions we are interested in are the following:
\begin{align}
    H_J = J \sum_i  \Vec{S}_X \cdot \Vec{S}_Y - V \sum_i n^X_i n^Y_i
\end{align}
We decompose these interactions in the pairing, spin and charge channels.
\begin{align}
    U_{abcd}^c = & \begin{pmatrix}
        0 & 0 &0 &-V\\
        0 & 0 &0 &0\\
        0 & 0 &0 &0\\
        -V & 0 &0 &0\\
    \end{pmatrix}\\
    U_{abcd}^z = & \begin{pmatrix}
        0 & 0 &0 &J\\
        0 & 0 &0 &0\\
        0 & 0 &0 &0\\
        J & 0 &0 &0\\
    \end{pmatrix}\\
\end{align}
The multiorbital RPA equation reads:
\begin{align}
    \chi^{RPA}(Q,0) = & \chi^0(Q,0) - \chi^0(Q,0) U_{\text{eff}} \chi^{RPA}\\
    \chi^{RPA} = & (1+\chi^0 U_{\text{eff}})^{-1} \chi^0
\end{align}
\\
Normal state instabilities are defined as divergences in $\chi^{RPA}$, or equivalently, $\text{max}_\lambda(\chi^0 U) = 1$. This signals an instability towards the order given by the corresponding eigenvector.
\\
In the case of singlet pairing, since we are only decomposing in the OTSS pairing channel, the effective interaction is a scalar and corresponds to a single element of the complete effective interaction tensor:
\begin{align}
    U_{XXYY}^p = \frac{-3}{4}J - V
\end{align}
\\
\section{Mean-Field theory}
We consider a local spin-spin interaction across different orbitals.\\
\begin{align}
    \sum_r J S_r^X S_r^Y =& \sum_r \frac{J}{4} X_\a^\dagger \s_{\a \b}^a X_\b Y_\g^\dagger \s_{\g \d}^a Y_\d \nonumber \\
     = & \frac{J}{4} \sum_r \left(2 X_\a^\dagger Y_\b^\dagger Y_\a X_\b - X_\a^\dagger Y_\b^\dagger Y_\b X_\a \right)
\end{align}

We define the following pairing order parameter:
\begin{equation}
    \Delta_r^* = \expval{X_{r \ua}^\dagger Y_{r\da}^\dagger}, -\Delta_r^* = \expval{X_{r \da}^\dagger Y_{r\ua}^\dagger},\expval{X_{r\a}^\dagger Y_{r\b}^\dagger} = i\s_{\a\b}^y \Delta_r^*, \expval{Y_{r\b} X_{r\a}} = i\s_{\a\b}^y\Delta_r
\end{equation}

Mean-field decoupling the interaction in the singlet channel yields the following:

\begin{align*}
    & \frac{J}{4}\sum_r \left[2\left(\expval{X_\a^\dagger Y_\b^\dagger} Y_\a X_\b + X_\a^\dagger Y_\b^\dagger\expval{Y_\a X_\b} - \expval{X_\a^\dagger Y_\b^\dagger} \expval{Y_\a X_\b}\right) - \expval{X_\a^\dagger Y_\b^\dagger} Y_\b X_\a - X_\a^\dagger Y_\b^\dagger \expval{Y_\b X_\a} + \expval{X_\a^\dagger Y_\b^\dagger} \expval{Y_\b X_\a} \right] \\ 
    &= \frac{-3J}{4} \sum_r\left[\Delta_r^* \left(Y_{\da}X_{\ua} - Y_{\ua}X_{\da}\right) + \Delta_r \left(X_{\ua}^\dagger Y_{\da}^\dagger - X_{\da}^\dagger Y_{\ua}^\dagger\right)\right] + 2\frac{3J}{4} \sum_r \Delta_r \Delta_r^* \\
    &= J_{\text{eff}} \sum_r\left[\Delta_r^* \left(Y_{\da}X_{\ua} - Y_{\ua}X_{\da}\right) + \Delta_r \left(X_{\ua}^\dagger Y_{\da}^\dagger - X_{\da}^\dagger Y_{\ua}^\dagger\right)\right] - 2\sum_r J_{\text{eff}} \Delta_r \Delta_r^*
\end{align*}

We now consider the Fourier transformed order parameter, and restrict the sum over Q, to a single momentum. More generally, coupling of all wave-vectors $Q$ should be considered. Resorting to a single $Q$ captures a Fulde-Ferrel type pair density wave.
\begin{align}
    \Delta_r^* & = \sum_{k,k'}\expval{X_{k\ua}^\dagger Y_{k'\da}^\dagger} e^{i(k+k') r} \nonumber \\
    &= \frac{1}{N} \sum_{k,Q} \expval{X_{k\ua}^\dagger Y_{-k+Q\da}^\dagger} e^{iQr} \nonumber\\
    &= \frac{1}{N} \sum_{k} \expval{X_{k\ua}^\dagger Y_{-k+Q\da}^\dagger} e^{iQr} \nonumber \\
    &=\Delta_0 e^{iQr}
\end{align}

We can now setup the BdG Hamiltonian for our system.
\begin{align*}
    H_0 & = \sum_k \left[(\e_k^X - \mu) X_{k\ua}^\dagger X_{k\ua} + (\e_k^Y -\mu)X_{Y\ua}^\dagger Y_{k\ua} + \e_{k}^{XY}( X_{k\ua}^\dagger Y_{k\ua} + + \text{h.c})\right. \\
    &\left. + (\e_{-k+Q}^X - \mu) X_{-k+Q\da}^\dagger X_{-k+Q\da} + (\e_{-k+Q}^Y -\mu)X_{-k+Q\da}^\dagger Y_{-k+Q\da} + \e_{-k+Q}^{XY} (X_{-k+Q\da}^\dagger Y_{-k+Q\da} + \text{h.c})\right]\\
\end{align*}
\begin{align}
    H_{BdG} = \sum_k  \begin{pmatrix}
X_{k\ua}^\dagger
& Y_{k\ua}^\dagger
& X_{-k+q\da}
& Y_{-k+q\da}
\end{pmatrix}
\begin{pmatrix}
    \e_k^X - \mu & \e_{k}^{XY} & 0 &J_{\text{eff}}\D_0 \\
    \e_{k}^{XY} & \e_k^Y - \mu & J_{\text{eff}}\D_0 & 0 \\
    0 & J_{\text{eff}}\D_0^* & -\e_{-k+Q}^X + \mu & -\e_{-k+Q}^{XY} \\
    J_{\text{eff}}\D_0^* & 0 & -\e_{-k+Q}^{XY} & -\e_{-k+Q}^Y + \mu
\end{pmatrix}
\begin{pmatrix}
X_{k\ua} \\ 
Y_{k\ua}\\ 
X_{-k+q\da}^\dagger \\
Y_{-k+q\da}^\dagger
\end{pmatrix}\\
 + \sum_k\left[\e_{-k+Q}^X - \mu + \e_{-k+Q}^Y - \mu \right] - 2\sum_r J_{\text{eff}}\D_0 \D_0^* \nonumber
\end{align}



\end{document}

%% file: defns.tex
\usepackage[normalem]{ulem}

\newcommand{\bea}{\begin{eqnarray}}
\newcommand{\eea}{\end{eqnarray}}

\newcommand{\be}{\begin{equation}}
\newcommand{\ee}{\end{equation}}
\newcommand{\bk}{{{\bf{k}}}}

\newcommand{\bQ}{{{\bf{Q}}}}

\newcommand{\beal}{\begin{align}}
\newcommand{\eeal}{\end{align}}
\newcommand{\ra}{\rangle}
\newcommand{\la}{\langle}

\newcommand{\upa}{\uparrow}
\newcommand{\dna}{\downarrow}

\newcommand{\dg}{{\dagger}}
\newcommand{\pdg}{{\phantom\dagger}}

\newcommand{\btjstrw}{\mathrel{{\rotatebox[origin=c]{90}
{$\bowtie$}}\kern-0.18em\raisebox{-.95ex}{$\bullet$}
\kern-0.5em\raisebox{.97ex}{$\bullet$}
\kern-1.12em\raisebox{.97ex}{$\bullet$}
\kern-0.52em\raisebox{-.95ex}{$\bullet$}}}

\newcommand{\btjnbrR}{{\mathrel{\rotatebox[origin=c]{90}
{$\bowtie$}}\kern-0.22em\raisebox{.9ex}{$\bullet$}
\kern-1.em\raisebox{-.8ex}{$\bullet$}}}
\newcommand{\btjnbrL}{{\mathrel{\rotatebox[origin=c]{90}
{$\bowtie$}}\kern-0.22em\raisebox{-.8ex}{$\bullet$}
\kern-1.em\raisebox{+.9ex}{$\bullet$}}}

\def\a{\alpha}
\def\b{\beta}

\def\d{\delta}
\def\e{\epsilon}

\def\g{\gamma}

\def\s{\sigma}
\def\t{\tau}

\def\w{\omega}

\def\D{\Delta}

\def\W{\Omega}

\newcommand{\llangle}[1][]{\savebox{\@brx}{\(\m@th{#1\langle}\)}%
  \mathopen{\copy\@brx\kern-0.5\wd\@brx\usebox{\@brx}}}
\newcommand{\rrangle}[1][]{\savebox{\@brx}{\(\m@th{#1\rangle}\)}%
  \mathclose{\copy\@brx\kern-0.5\wd\@brx\usebox{\@brx}}}

\def\ua{\uparrow}
\def\da{\downarrow}